\documentclass[journal]{IEEEtran}

\ifCLASSINFOpdf
\else
\fi

\usepackage{cite}
\usepackage{graphicx}
\usepackage{caption}
\usepackage{physics, amsmath,amssymb,amsfonts,algpseudocode}
\usepackage{upgreek}
\usepackage{optidef}
\usepackage{multirow}
\usepackage{array}
\usepackage{stfloats}
\usepackage{amsthm}
\usepackage{lipsum}
\usepackage{mathtools}
\usepackage{cuted}
\usepackage{algorithm}
\usepackage{blindtext}
\usepackage{amsfonts}
\usepackage{gensymb}

\usepackage{textcomp}
\usepackage{todonotes}
\begin{document}

\title{Deep Reinforcement Learning based Joint Spectrum Allocation and Configuration Design for STAR-RIS-Assisted V2X Communications}

\author{Pyae~Sone~Aung,
        Loc~X.~Nguyen,
        Yan~Kyaw~Tun,~\IEEEmembership{Member,~IEEE,}
        Zhu~Han,~\IEEEmembership{Fellow,~IEEE,}
        and~Choong~Seon~Hong,~\IEEEmembership{Senior Member,~IEEE,}
\thanks{Pyae Sone Aung, Loc X. Nguyen and Choong Seon Hong are with the
Department of Computer Science and Engineering, Kyung Hee University, Yongin-si, Gyeonggi-do 17104, Rep. of Korea, e-mail:\{pyaesoneaung, xuanloc088, cshong\}@khu.ac.kr.}
\thanks{Yan Kyaw Tun is with Division of Network and Systems Engineering, School of Electrical Engineering and Computer Science, KTH Royal Institute of Technology, Brinellvägen 8, 114 28 Stockholm, Sweden, e-mail:{\{yktun\}@kth.se}.}
\thanks{Zhu Han is with the Electrical and Computer Engineering Department, University of Houston, Houston, TX 77004, and the Department of Computer Science and Engineering, Kyung Hee University, Yongin-si, Gyeonggi-do 17104, Rep. of Korea, email\{hanzhu22\}@gmail.com.}}

\maketitle

\begin{abstract}
Vehicle-to-Everything (V2X) communications play a crucial role in ensuring safe and efficient modern transportation systems. However, challenges arise in scenarios with buildings, leading to signal obstruction and coverage limitations. To alleviate these challenges, reconfigurable intelligent surface (RIS) is regarded as an effective solution for communication performance by tuning passive signal reflection. RIS has acquired prominence in 6G networks due to its improved spectral efficiency, simple deployment, and cost-effectiveness. Nevertheless, conventional RIS solutions have coverage limitations. Therefore, researchers have started focusing on the promising concept of simultaneously transmitting and reflecting RIS (STAR-RIS), which provides 360\degree coverage while utilizing the advantages of RIS technology. In this paper, a STAR-RIS-assisted V2X communication system is investigated. An optimization problem is formulated to maximize the achievable data rate for vehicle-to-infrastructure (V2I) users while satisfying the latency and reliability requirements of vehicle-to-vehicle (V2V) pairs by jointly optimizing the spectrum allocation, amplitudes, and phase shifts of STAR-RIS elements, digital beamforming vectors for V2I links, and transmit power for V2V pairs. Since it is challenging to solve in polynomial time, we decompose our problem into two sub-problems. For the first sub-problem, we model the control variables as a Markov Decision Process (MDP) and propose a combined double deep Q-network (DDQN) with an attention mechanism so that the model can potentially focus on relevant inputs. For the latter, a standard optimization-based approach is implemented to provide a real-time solution, reducing computational costs. Extensive numerical analysis is developed to demonstrate the superiority of our proposed algorithm compared to benchmark schemes.
\end{abstract}


%
\IEEEpeerreviewmaketitle

\section{Introduction}


\subsection{Background and Motivations}
In recent years, Vehicle-to-Everything (V2X) communications, which enable communication between vehicles and other entities in surrounding environments, hold significant potentials for improving road safety, traffic efficiency, and overall driving experience. V2X includes several communication modes, including V2I (Vehicle-to-Infrastructure) and V2V (Vehicle-to-Vehicle). Each communication mode has its unique characteristics and requirements. The V2I mode enables vehicles to communicate with infrastructure elements, such as traffic lights, road signs, and roadside sensors. Furthermore, the V2I mode can provide real-time traffic information, road conditions, and traffic signals to vehicles, which can enhance safety and reduce traffic congestion. V2I mode can also enable the infrastructure to deliver infotainment content such as video streaming and crowd-sensing \cite{abboud2016interworking}. The V2V mode enables vehicles to communicate with other vehicles in their vicinity. It can distribute information on the position, speed, and direction of nearby vehicles. Moreover, the V2V mode can also enable vehicles to form ad-hoc networks, which can improve communication reliability. Nonetheless, V2X communications require reliable and robust connectivity between vehicles and other entities in the environment. Therefore, signal blockage, interference, and coverage limitations can potentially degrade the performance of V2X communications significantly.

To address these challenges, reconfigurable intelligent surface (RIS), which consists of passive and reconfigurable elements, has emerged as a promising technology. An RIS is a meta-surface consisting of a planar array of passively reflecting elements that manipulate wireless signals in the desired direction through customized phase response \cite{huang2019reconfigurable}. By reflecting and rerouting wireless signals, RIS can improve signal coverage, reduce interference, and increase the capacity and efficiency of V2X communication. Furthermore, the increasing popularity of RISs may be due to the fact that they are cost-effective and simple to deploy for improving spectral efficiency. In contrast to more conventional forms of collaborative communication such as decode-and-forward (DF) and amplify-and-forward (AF), RIS does not call for the use of an extra power amplifier. As a result, it is more sustainable to the environment and utilizes less resources \cite{liu2021reconfigurable}. Nevertheless, despite these benefits, conventional RIS has a number of limitations. One main technical challenge of conventional RIS is that since it only supports reflecting, the transmitter and receiver need to be on the same side, which can restrict the flexibility and practical deployment of RIS in various wireless communication applications.

To overcome this limitation, simultaneously transmitting and reflecting RIS (STAR-RIS) offers a potential solution. As the name implies, STAR-RIS can reflect the incident signal on the same side, while transmitting the incident signal on the opposite side. The transmitted and reflected signals can be adjusted through two generally independent coefficients, namely the transmission and the reflection coefficients, by altering the electric and magnetic currents of a STAR-RIS element. Therefore, STAR-RIS may provide more degrees of freedom (DoFs) for manipulating signal propagation, leading to 360$\degree$ coverage. In the V2X communication environment, STAR-RIS can be installed on buildings, road signs, or traffic posts. STAR-RIS in V2X communication can lead to extended communication range, improved reliability, enhanced interference mitigation, optimized channel characteristics, precise localization, and increased energy efficiency \cite{guo2023star}. These benefits can significantly contribute to the development of safer, more efficient, and intelligent V2X systems.

\subsection{Challenges and Research Contributions}
V2X communications encounter several challenges. For instance, in areas with tall buildings and obstacles, wireless connections between vehicular user equipments (VUEs) and base station (BS) may be vulnerable. Hence, the service quality falls below the desirable levels. Moreover, V2V links become less reliable when the blockage effect is considered, which restricts the performance of V2V communications. There have been several works assisting with conventional RIS in vehicular networks to improve the spectral efficiency between the vehicles and BS \cite{al2022reconfigurable, chen2022reconfigurable}. However, with RIS, due to its limitation, vehicles and BS need to be on the same side, which hinders the DoFs. In this paper, to satisfy the distinct QoS requirements for both V2I and V2V communications without the requirement of external power sources,  we propose STAR-RIS-assisted V2X communications. The contributions of the paper are organized as follows:
\begin{itemize}    
    \item Firstly, we propose a STAR-RIS-assisted V2X communication system and formulate the optimization problem to maximize the achievable data rates of VUEs while satisfying the reliability and latency requirements of V2V pairs by jointly optimizing the spectrum allocation, amplitude, and phase shift of STAR-RIS elements, digital beamforming vector towards VUEs, and power control for cellular V2X communications.
    \item Secondly, since the formulated problem is mixed-integer and non-linear, it is challenging to solve in polynomial time. Therefore, we decompose our problem into two sub-problems. Then, for the first sub-problem, we propose a DRL-based algorithm where we combine the deep residual network (Resnet) with attention layers in DDQN to enable the model to focus on particular information, thus reducing redundancy. For the second sub-problem, we employ a standard optimization approach to ensure its efficiency and reduce computational costs. We solve the sub-problems alternately until we reach convergence.
    \item Finally, a comprehensive numerical analysis is conducted to demonstrate the effectiveness of our proposed system in comparison to the conventional RIS-assisted V2X communication system. Moreover, our proposed solution approach achieves greater performance compared to vanilla DDQN, ensuring computational efficiency and reducing the computational burden for real-world implementation.
\end{itemize}

The rest of the paper is categorized as follows: we present the related works in Section \ref{relatedworks}. Next, we present our system model and problem formulation in Section \ref{systemmodel}. Afterward, the solution approach is proposed in Section \ref{solution}, and performance evaluation is performed in Section \ref{evaluation}. Finally, Section \ref{conclusion} concludes our paper.
\section{Related Works}\label{relatedworks}
\subsection{V2X Communications}
An overview on the literature related to V2X communications is discussed in this sub-section \cite{ye2019deep, zhang2019deep, banitalebi2022distributed, xu2022learning, tian2023deep, chen2018interference, zhang2021centralized}. The authors in \cite{ye2019deep} studied the DRL-based decentralized resource allocation system for V2V communications in which each V2V link is considered as an agent and everything outside the specific V2V link is treated as the environment. The authors in \cite{zhang2019deep} investigated to satisfy the latency and reliability requirements of V2V pairings while maximizing the combined capacity of vehicle-to-infrastructure users. The authors in \cite{banitalebi2022distributed} proposed to improve energy efficiency by jointly optimizing the sub-carrier and power allocation in a distributed manner for the cellular V2X (C-V2X) network. The authors in \cite{xu2022learning} examined the air slicing-enabled unmanned aerial vehicles (UAVs)-assisted C-V2X communication system with the aim of improving bandwidth efficiency while simultaneously ensuring the transmission rate and the latency in delivering service to automotive customers. The authors in \cite{tian2023deep} evaluated the DRL-based resource allocation to fulfill the heterogeneous QoS requirements for C-V2X communication in a vehicular network. The authors in \cite{chen2018interference} proposed the spatial reuse-based resource allocation for different device-to-device-enhanced V2X communication in a non-orthogonal multiple access (NOMA)-integrated system. Moreover, in \cite{zhang2021centralized}, the authors investigated the resource allocation with the graph-based matching approach to improve system capacity for new radio V2X communications in the NOMA scheme. However, the previous works did not consider STAR-RIS, with which the coverage and range can be extended along with improved signal quality and reliability.

\subsection{RIS-assisted Communications}
An overview of the literature related to RIS-assisted communications is discussed in this sub-section \cite{aung2022energy, aung2022energy1, chen2020reconfigurable, di2020hybrid, zhang2020reconfigurable}. For our previous works, in \cite{aung2022energy}, we studied the DRL-based energy-efficient downlink communication system for multiple RISs-enabled networks to provide robust transmission and establish multiple paths to enhance the received signal strength, and in \cite{aung2022energy1} we elevated to multiple aerial RISs (ARISs) networks where RISs are implemented on UAVs to provide better line-of-sight and improved spectral efficiency between the BS and user equipments (UEs). The authors in \cite{chen2020reconfigurable} proposed to maximize the system sum rate in RIS-assisted device-to-device (D2D) communication in heterogeneous cellular networks. In \cite{di2020hybrid}, the authors investigated the hybrid beamforming scheme between passive beamforming and active beamforming for sum rate maximization in a downlink RIS-based multi-user system. The authors in \cite{zhang2020reconfigurable} studied the impact of the number of phase shifts on the effectiveness of the achievable RIS-assisted uplink communication system.

Additionally, there are various studies on RIS-assisted in V2X communications\cite{saikia2023proximal, al2022reconfigurable, bansal2022rate, chen2022reconfigurable}. In \cite{saikia2023proximal}, the authors proposed to maximize the achievable data rate of RIS-assisted full duplex vehicular communications with a low complexity proximal policy optimization (PPO) algorithm. The authors in \cite{al2022reconfigurable} analyzed a DRL-based vehicular communication network in which RIS is utilized to establish communication links between the road side unit (RSU) and the vehicles that are moving via out-of-service zones. The authors in \cite{bansal2022rate} investigated the rate-splitting multiple access scheme for the RIS-enabled UAV-based vehicular communication network to minimize the average multi-user outage probability. In \cite{chen2022reconfigurable}, the authors studied protocols and performance on resource allocation in RIS-aided vehicular networks under statistical CSI and instantaneous CSI, respectively. However, the aforementioned works did not take into account reliability and latency for the V2V users, which also play a crucial role in V2X communication, in addition to not applying STAR-RIS, which restricts coverage expansion.

\subsection{STAR-RIS-assisted Communications}
An overview of the literature related to STAR-RIS-assisted communications is discussed in this sub-section \cite{wu2022resource, qin2023joint, guo2023star, chen2022robust}. In \cite{guo2023star}, the authors investigated the outage behavior in the STAR-RIS-empowered cognitive non-terrestrial vehicular network in the NOMA scheme. The authors in \cite{wu2022resource} discussed the resource allocation in STAR-RIS-assisted multi-carrier in both orthogonal multiple access (OMA) and NOMA schemes. In \cite{qin2023joint}, the authors investigated maximizing the computation rate of all users in energy-efficient wireless-powered mobile edge computing (MEC) systems. There exist several works on STAR-RIS-assisted vehicular networks. In order to improve transmission for users within the vehicle as well as for users of nearby vehicles, the authors of \cite{chen2022robust} proposed the installation of the active reconfigurable intelligent omni-surface (RIOS), which is similar to STAR-RIS, on the window of the vehicle.

The aforementioned studies did not take into account STAR-RIS in V2X communication networks, which required coping with challenges imposed by different QoS requirements for V2I users and unreliable V2V links. Our objective is to maximize the spectral efficiency for V2I users and meet the requirements for reliability and latency for V2V pairs. To address the aforementioned challenges, a unique solution approach based on the DDQN-attention method and an efficient optimization technique is proposed.
\\ 

\section{System Model}\label{systemmodel}
\begin{figure}[t]
	\includegraphics[width=\linewidth]{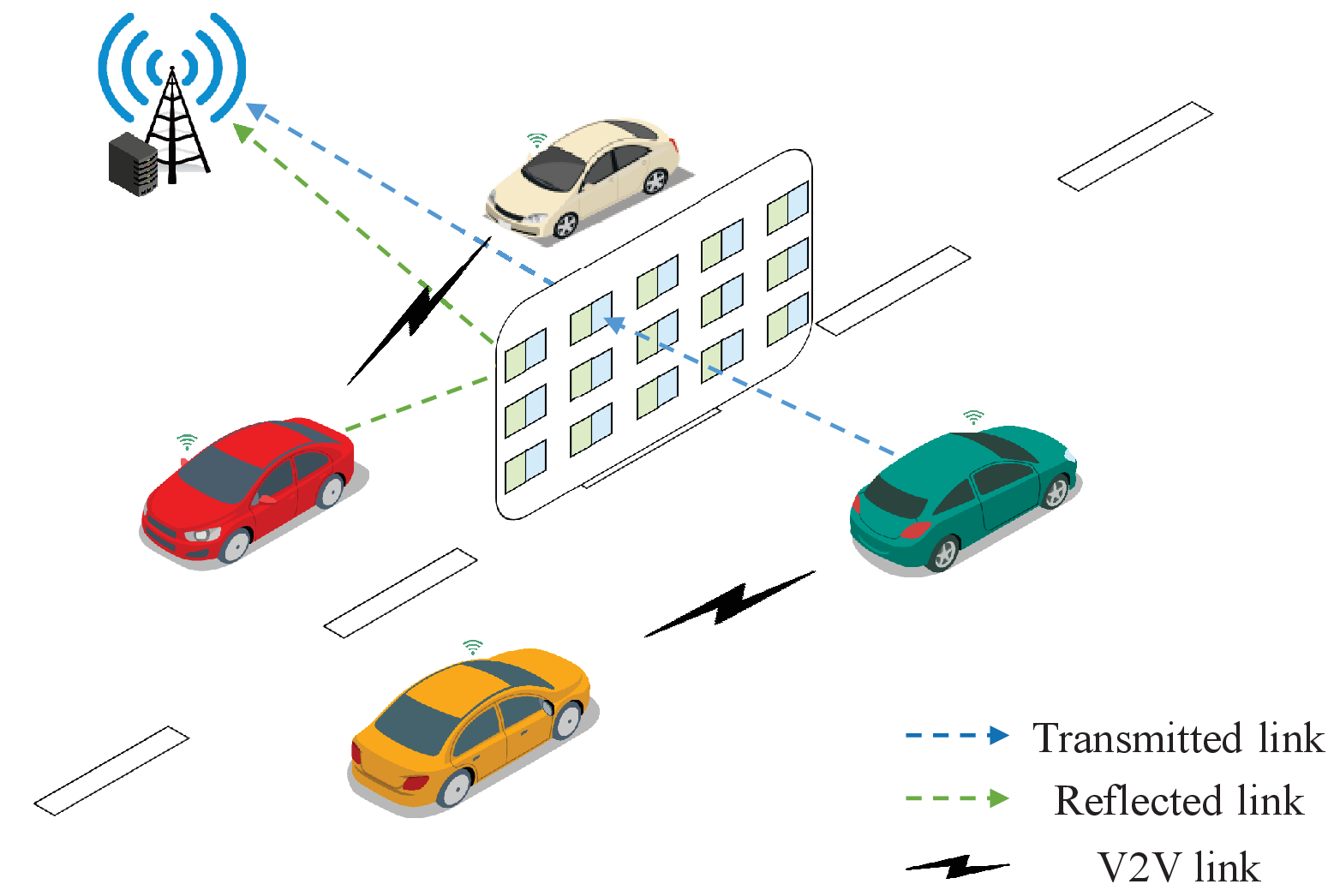}
	\caption{System model of STAR-RIS assisted V2X communications.}
	\label{sm}
\end{figure}

As shown in Fig.~\ref{sm}, we consider a STAR-RIS assisted V2X communication environment, which consists of a single BS with multiple antennas $\mathcal{B} = \{1, 2, \dots, B\}$ and a set $\mathcal{I}$ of $I$ single-antenna vehicular user equipments (VUEs), and a set $\mathcal{V}$ of $V$ V2V user pairs, respectively. The VUEs upload bandwidth-intensive contents such as videos, and photos to the BS, assisted by STAR-RIS due to the weak LoS link between the BS and vehicles. Simultaneously, each V2V pair has one V2V transmitter and one V2V receiver for the delivery of safety-critical messages between the vehicles. In our scenario, the same VUE-equipped vehicle functions as a V2V user simultaneously as shown in Fig. \ref{sm}. Furthermore, the STAR-RIS consists of an array of $\mathcal{N} = \{1, 2, \dots, N\}$ elements and the BS controls the transmission and reflection coefficient via a separate wireless link. Moreover, as in \cite{zhai2019joint}, we assume that V2I channels are orthogonally allocated and each VUE is assigned a single channel. To enhance spectrum usage efficiency, the V2V pair can reuse the same spectrum with the VUE.

\textbf{Spectrum Allocation.} Therefore, we denote $\mathbf{A} \in \mathbb{R}^{|\mathcal{I}| \times |\mathcal{V}|}$ as the spectrum allocation matrix for all V2V pairs over all spectrum allocated for them. Then, the spectrum allocation variable for each V2V pair $v$ can be defined as follows:
\begin{equation}
    a_{i,v} =\left \{ \begin{array}{ll}{1,} & {\text {if V2V pair $v$ reuses the same spectrum with}} \\ & {\text {$i$-th VUE,}} \\ {0,} & {\text {otherwise.}}\end{array}\right. 
\end{equation}


\subsection{V2I Channel Model}
We consider there exist both direct and indirect wireless communication links between VUEs and the BS for uplink communication. We assume that perfect channel state information (CSI) can be achieved by using the technique as in \cite{wu2021channel}. For the direct link, due to the lack of LoS links and potential extensive scattering, we adopt Rayleigh fading \cite{gu2022socially}. Therefore, the channel gain between the BS and the $i$-th VUE can be expressed as follows:
\begin{equation}
   \mathbf{h}_{i,B} = \sqrt{\eta d_{i,B}^{-\delta}}\tilde{h} \in \mathbb{C}^{1 \times |\mathcal{B}|},
\end{equation}
where $\delta \geq 2$ is the path loss exponent, $\eta$ is the channel gain at $1$ m reference distance, $d_{i,B}$ is the distance between the BS and the $i$-th VUE, and $\tilde{h}$ is the complex Gaussian random scattering component with the value, $\mathcal{C}\sim(0,1)$.

Since the indirect link is assisted by the STAR-RIS, we first define the reflection coefficient and transmission coefficient of $n$-element of STAR-RIS, ${\phi}^n_r$ and ${\phi}^n_t$ as follows:
\begin{equation}
    {\phi}^n_r = {\beta}^n_r e^{j{\theta}^n_r},
\end{equation}
\begin{equation}
    {\phi}^n_t = {\beta}^n_t e^{j{\theta}^n_t},
\end{equation}
where $\beta_r, \beta_t \in [0,1]$ denotes the amplitude of the reflection and transmission, and ${\theta}^n_r, {\theta}^n_t \in [0,2\pi)$ denotes the phase shift values of $n$-element, respectively. For simplicity, we assume that the phase-shift values are discrete, and phase-shift coefficients of the reflection and transmission of $n$-element can be independently adjusted, with the feasible range of phase shift values specified by
\begin{equation}
    {\theta}^n_\lambda = \frac{\kappa \pi}{2^{b-1}}, \lambda \in \{r,t\},
\end{equation}
where $b$ is the phase shift resolution in bits, and $\kappa \in \{0,1,\dots,2^b-1\}$ is the phase shift index \cite{zhao2022simultaneously}. From there, we can obtain the reflection and transmission coefficient matrices as follows:
\begin{equation}
    \mathbf{\Phi}_r = \operatorname{diag}({\beta}^1_r e^{j{\theta}^1_r}, {\beta}^2_r e^{j{\theta}^2_r}, \dots, {\beta}^N_r e^{j{\theta}^N_r}) \in \mathbb{C}^{|\mathcal{N}| \times |\mathcal{N}|},
\end{equation}
\begin{equation}
    \mathbf{\Phi}_t = \operatorname{diag}({\beta}^1_t e^{j{\theta}^1_t}, {\beta}^2_t e^{j{\theta}^2_t}, \dots, {\beta}^N_t e^{j{\theta}^N_t}) \in \mathbb{C}^{|\mathcal{N}| \times |\mathcal{N}|}.
\end{equation}
The indirect link consists of two components: V-STAR link (VUE to STAR-RIS) and the reflected/transmitted (STAR-RIS to BS) link. We adopt the Rician fading for both of these components \cite{al2022reconfigurable}. Consequently, the channel gain for the V-STAR link can be expressed as follows:
\begin{equation}
    \mathbf{h}_{i,N} = \sqrt{\eta d_{i,N}^{-\delta}} \sqrt{\frac{K}{1+K}}\mathbf{h}_{i,N}^{\mathrm{LoS}} + \sqrt{\frac{1}{1+K}}\mathbf{h}_{i,N}^{\mathrm{NLOS}} \in \mathbb{C}^{1 \times |\mathcal{N}|},
\end{equation}
where $K$ is the Rician factor, and $ d_{i,N}$ is the distance between the $i$-th VUE and STAR-RIS, $\mathbf{h}_{i,N}^{\mathrm{LoS}}$ is the deterministic LoS component between the $i$-th VUE and STAR-RIS, as determined by the azimuth angle-of-arrival (AoA) of the link, and $\mathbf{h}_{i,N}^{\mathrm{NLOS}}$ is the non-LoS component with a circularly-symmetric i.i.d. complex Gaussian distribution, respectively \cite{aung2023deep}. Similarly, the channel gain for the reflected/transmitted link can be represented as
\begin{equation}
\mathbf{H}_{N,B} = \sqrt{\eta d_{N,B}^{-\delta}} \sqrt{\frac{K}{1+K}}\mathbf{H}_{N,B}^{\mathrm{LoS}} \in \mathbb{C}^{|\mathcal{N}| \times |\mathcal{B}|},
\end{equation}
where $d_{N,B}$ is the distance between STAR-RIS and the BS, and $\mathbf{H}_{N,B}^{\mathrm{LoS}}$ is the deterministic LoS component between the STAR-RIS and BS, as determined by the azimuth angle-of-departure (AoD) of the link. Therefore, the channel model between VUE $i$ and BS can be obtained as
\begin{equation}
    \mathbf{h}_i = \mathbf{h}_{i,B} + \mathbf{h}_{i,N}\mathbf{\Phi}_{\lambda} \mathbf{H}_{N,B}, \lambda \in \{r,t\}.
\end{equation}
Hence, the received signal at the BS can be achieved as
\begin{equation}
    \mathbf{y} = \sum_{i\in\mathcal{I}} \mathbf{h}_i\mathbf{x}_i + \boldsymbol{\omega},
\end{equation}
where $\mathbf{x}_i= \mathbf{p}_i s$ is the transmitted signal with the digital beamforming vector $\mathbf{p}_i$ for the $i$-th VUE and unit-power information symbol $s$ \cite{huang2019reconfigurable}, and $\boldsymbol{\omega}\sim\mathcal{CN}(\mathbf{0},\sigma^2\mathbf{I}_B)$ is the additive white Gaussian noise (AWGN) at the BS with zero mean and variance $\sigma^2$. Afterwards, the uplink signal-to-interference-plus-noise ratio (SINR) at the $i$-th VUE can be expressed as
\begin{equation}
    \gamma_i = \frac{|\mathbf{h}_i\mathbf{p}_i|^2}{G_{v}+\sigma^2},
\end{equation}
where $G_v$ is the interference power of the V2V pair which shares the same spectrum with V2I link and have the formula as
\begin{equation}
   G_{v} = \sum_{v \in \mathcal{V}}\left(a_{i,v} p_v |h_v|^2\right),
\end{equation}
where $p_v$ is the transmit power of V2V transmitter for the $v$-th V2V pair, and $h_v$ is the channel gains between the V2V transmitter to the BS, respectively. Consequently, we can get the achievable data rate of the $i$-th VUE can be obtained as
\begin{equation}
    R_i =  W_0 \log_{2}(1+\gamma_i),
\end{equation}
where $W_0$ is the transmission bandwidth.

\subsection{V2V Communication Model}
 In V2V mode, the communication links such as the delivery of safety-critical messages are directly transferred between two devices via V2V transmitter and V2V receiver. The signal-to-interference-plus-noise ratio (SINR) of the $v$-th V2V pair can be represented as
 \begin{equation}
     \gamma_{v} = \frac{p_v |h_v|^2}{G_i + \sigma^2},
 \end{equation}
where $G_i$ is the interference power from the V2I link sharing the same spectrum with the V2V link and is written as
\begin{equation}
    G_i = \sum_{i \in \mathcal{I}} \left(a_{i,v} |\mathbf{h}_i \mathbf{p}_i|^2\right).
\end{equation}
Similarly, the achievable data rate of V2V pair $v$ can be obtained as
\begin{equation}
    R_v = W_0 \log_{2}(1 + \gamma_v).
\end{equation}

\subsection{Problem Formulation}
In the V2X communication system, each sort of communication requires distinct QoS requirements. V2I communication demands extensive bandwidth for infotainment services whilst V2V communication depends on low latency and reliability for the transmission of emergency alerts and traffic information messages. We initially determine a set of constraints and limitations for our STAR-RIS-assisted V2X communication system, which are as follows.

Firstly, we need to consider the minimum achievable data rate for the $i$-th VUE for quality of service (QoS) requirement as
\begin{equation}
    R_i \geq R_{\min}, \forall i \in \mathcal{I},
\end{equation}
where $R_{\min}$ is the minimum data rate specified for each VUE. Next, we also need to consider the latency requirements for the delivery of payload of size $D$ within time limits for the $v$-th V2V pair and can be written as
\begin{equation}
    R_v \geq \frac{D}{T_{\max}}, \forall v \in \mathcal{V},
\end{equation}
where $D$ is the size of the data payload in bits and $T_{\max}$ is the maximum time budget required for the delivery of the payload between the V2V pair. For reliability, the outage probability is considered as a reliability constraint for the $v$-th V2V pair can be expressed as
\begin{equation}
    \mathbb{P}\{\gamma_v \leq \gamma_0\} \leq p_0, \forall v \in \mathcal{V},
\end{equation}
where $\gamma_0$ and $p_0$ are the outage threshold and outage probability, respectively. According to \cite{zhang2019deep}, this term can be converted into the following as
\begin{equation}
    \gamma_v \leq \gamma_{\mathrm{ef}} = \frac{\gamma_0}{\ln{\left(\frac{1}{1-p_0}\right)}}, \forall v \in \mathcal{V},
\end{equation}
where $\gamma_{\mathrm{ef}}$ is the effective outage threshold. We assume all V2V pairs have the same payload size, time budget, and outage probability, respectively. To minimize cross-layer interference, each V2V pair is only permitted to access the spectrum of one V2I link, while the spectrum of each V2I link can be allocated to only one V2V pair. To satisfy this, we add the following constraints as 
\begin{equation}
    a_{i,v} \in \{0, 1\}, \forall i \in \mathcal{I},
\end{equation}
\begin{equation}
    \sum_{v \in \mathcal{V}} a_{i,v} \leq 1, \forall i \in \mathcal{I}.
\end{equation}

For STAR-RIS, the accessible phase shifts for the reflective/transmissive elements are as in the range as
\begin{equation}
    0 \leq \theta^n_r,\ \ \theta^n_t < 2\pi, \forall n \in \mathcal{N}.
\end{equation}
Next, we implement the energy-splitting protocol for the operation of STAR-RIS, in which the energy of the incident signal is divided into the energies of the transmitted and reflected signals \cite{mu2021simultaneously}. We assume there is no energy dissipation by STAR-RIS and hence, due to the law of conservation of energy, we have
\begin{equation}
    (\beta_r)^2 + (\beta_t)^2 = 1.
\end{equation}

Considering these, the main objective of our STAR-RIS-assisted V2X communication system is to maximize the average achievable data rate for the VUEs while satisfying the reliability and latency requirements for the V2V pairs. Therefore, the optimization problem can be formulated as
\begin{maxi!}|s|
	{\mathbf{A}, \boldsymbol{\beta}_\lambda, \boldsymbol{\theta}_\lambda, \mathbf{p}_\mathrm{V2I}, \mathbf{p}_\mathrm{V2V}} {\sum_{i \in \mathcal{I}}R_i}
	{\label{OF}}{}\label{OFa}
	\addConstraint{R_i \geq R_{\min}, \forall i \in \mathcal{I}}\label{c1}
	\addConstraint{R_v \geq \frac{D}{T_{\max}}, \forall v \in \mathcal{V}}\label{c2}
	\addConstraint{\gamma_v \leq \gamma_{\mathrm{ef}} = \frac{\gamma_0}{\ln{\left(\frac{1}{1-p_0}\right)}}, \forall v \in \mathcal{V}}\label{c3}
    \addConstraint{a_{i,v} \in \{0, 1\}, \forall i \in \mathcal{I}}\label{c5}
    \addConstraint{\sum_{v \in \mathcal{V}} a_{i,v} \leq 1, \forall i \in \mathcal{I}}\label{c6}
	\addConstraint{0 \leq \theta^n_r, \theta^n_t < 2\pi, \forall n \in \mathcal{N}}\label{c4}
    \addConstraint{(\beta_r)^2 + (\beta_t)^2 = 1}\label{cbeta}
    \addConstraint{\mathbf{p}_\mathrm{V2I}^H \mathbf{p}_\mathrm{V2I} \leq P_{\mathrm{I}\_{\max}}}\label{c7}
    \addConstraint{p_v \leq P_{\mathrm{V}\_{\max}}, \forall v \in \mathcal{V},}\label{c8}
\end{maxi!}
where $\boldsymbol{\beta}_\lambda = \{\beta^1_\lambda, \dots, \beta^N_\lambda\}$, $\lambda \in \{r, t\}$ is the amplitude vector, $\boldsymbol{\theta}_\lambda = \{\theta^1_\lambda, \dots, \theta^N_\lambda\}$, $\lambda \in \{r, t\}$ is the phase shift vector of STAR-RIS elements, $\mathbf{p}_\mathrm{V2I} = \{\mathbf{p_1}; \dots; \mathbf{p_I}\}$ is the digital beamforming vectors for VUEs, and $\mathbf{p}_\mathrm{V2V} = \{p_1, \dots, p_V\}$ is the transmit power vector from V2V transmitter between each V2V pair. Constraints (\ref{c7}) and (\ref{c8}) represent the maximum transmit power budget for V2I and V2V communications, respectively. The problem in (\ref{OF}) is a mixed-integer and non-linear programming (MINLP) problem due to binary control variables in (\ref{c5}) and non-convex constraints in (\ref{c1}), (\ref{c2}), (\ref{c3}), and (\ref{c7}), and coupling between decision variables in both objective function and constraints. Hence, there exist numerous local optimal solutions and it is challenging to solve in polynomial time. To tackle this type of problem, we first decompose the formulated problem into two sub-problems: joint spectrum allocation, amplitude, phase shift and transmit power problem, and digital beamforming problem. Then, we alternately solve the sub-problems until convergence is reached. For the first sub-problem, we propose DRL as a solution approach since the environment in V2X communications is changing and inconsistent, with a large number of possible actions and uncertain outcomes. For the second sub-problem, we implement the convex optimization method. The reason we do not apply DRL for the entire optimization problem because the action spaces combined for all spectrum allocation, amplitudes, and phase shift coefficients of STAR-RIS elements, digital beamforming for VUEs, and transmit power from V2V transmitter will be too large and make DRL agent difficult to learn. Similar to \cite{al2022reconfigurable} and \cite{lee2020deep}, it is practical to solve one decision to optimization approach while committing the other decisions to the machine learning approach to reduce the computational cost.\footnote{Due to the process of decoupling, there will be an unavoidable decrease in performance as the inherent complexity of the problem and the presence of unidentified parameters prevent achieving optimal solutions.
Among the several DRL algorithms, we employ DDQN with attention and the motivation and detailed discussion for DDQN-attention are elaborated on in the next section.}

\section{Solution Approach}\label{solution}
In order to address the MINLP problem, we first decompose our optimization problem into two-subproblems: 1) Joint spectrum allocation, amplitude, phase shift, and transmit power problem; and 2) digital beamforming problem. After a brief overview of DRL in Section $\ref{overviewdrl}$, we propose a DDQN-attention approach to solve the first sub-problem in Section $\ref{DA-STAR}$, and a standard optimization approach to solve the second sub-problem in Section $\ref{dbp}$.
\subsection{Overview of DRL}\label{overviewdrl}
DRL is a sub-field of machine learning that combines reinforcement learning algorithms with deep neural networks to learn complex decision-making policies. DRL algorithms involve training an artificial agent to learn optimal actions in an environment based on rewards and penalties received as feedback for the actions taken. These algorithms use deep neural networks to approximate the value function or policy function that maps state-action pairs to the expected future rewards.

One of the popular classes of DRL algorithms is the Deep Q-Network (DQN), which uses a deep neural network to approximate the optimal action-value function. However, DQN suffers from overestimation bias, where the estimated Q-values can be inaccurate and lead to sub-optimal policies. To address this issue, Double Deep Q-Network (DDQN) was introduced, which uses two separate neural networks to select and evaluate actions. DDQN has been shown to reduce overestimation bias and lead to more stable and efficient training of the network \cite{van2016deep}.

\subsection{DDQN-attention based joint spectrum allocation, amplitude, phase shift and transmit power problem}\label{DA-STAR}
For the given digital beamforming vectors $\mathbf{p}_\mathrm{V2I}$, the first sub-problem can be expressed as
\begin{maxi!}|s|
	{\mathbf{A}, \boldsymbol{\beta}_\lambda, \boldsymbol{\theta}_\lambda, \mathbf{p}_\mathrm{V2V}} {\sum_{i \in \mathcal{I}}R_i}
	{\label{OFF1}}{}\label{OFaa1}
	\addConstraint{R_i \geq R_{\min}, \forall i \in \mathcal{I}}\label{cc11}
	\addConstraint{R_v \geq \frac{D}{T_{\max}}, \forall v \in \mathcal{V}}\label{cc22}
	\addConstraint{\gamma_v \leq \gamma_{\mathrm{ef}} = \frac{\gamma_0}{\ln{\left(\frac{1}{1-p_0}\right)}}, \forall v \in \mathcal{V}}\label{cc33}
    \addConstraint{a_{i,v} \in \{0, 1\}, \forall i \in \mathcal{I}}\label{cc55}
    \addConstraint{\sum_{v \in \mathcal{V}} a_{i,v} \leq 1, \forall i \in \mathcal{I}}\label{cc66}
	\addConstraint{0 \leq \theta^n_r, \theta^n_t < 2\pi, \forall n \in \mathcal{N}}\label{cc44}
    \addConstraint{(\beta_r)^2 + (\beta_t)^2 = 1.}\label{cbetaaa}
\end{maxi!}

Considering DRL is a flexible approach because it can handle complex, high-dimensional optimization problems with non-linear and non-convex constraints, it is best suited for a STAR-RIS-assisted V2X communication system where the environment can be unpredictable and dynamic, and there is a need to make effective use of limited communication resources. Since it can learn from interactions and make decisions about changes in the environment in real-time, the use of DRL in our proposed system can result in improved communication performance compared to traditional methods.

 Since DRL is perceived as Markov Decision Process (MDP), we must first define MDP as a 4-tuple consisting of states $\mathcal{S}$, actions $\mathcal{A}$, immediate reward $\mathcal{R}$ and transition function $\mathcal{P}$. At each time step, the policy function $\pi(a|s)$ directs the agent to select an action based on its current state and receive a reward signal from the environment. A stochastic policy $\pi(a|s)=\mathrm{Pr}(A_t=a|S_t=s)$ can be defined as the probability of taking action $a$ in state $s$ at time step $t$. The goal of the agent is to learn a policy that maximizes the expected cumulative reward function over time, which can be represented as
 \begin{equation}
     \max_\pi \mathbb{E}\left(\sum_{j=t}^{\infty} \zeta^{j-t}\mathcal{R}_j\right),
 \end{equation}
 where $\zeta$ is the discount factor to balance immediate and future rewards. We model our STAR-RIS-assisted V2X communication system as an MDP.
 \subsubsection{Agent} In our system model, the BS acts as an agent, which is the decision-maker. It selects actions based on its current state and receives rewards from the environment in accordance with its actions. 
 \subsubsection{Environment} The environment is the system that the agent interacts with. It consists of the BS, STAR-RIS, VUEs, V2V pairs, and channel models. The environment changes over time based on the actions taken by the agent.
 \subsubsection{States} The state of the system at any given time represents all the information that the agent needs to make a decision. States can either be fully observable or partially observable. In our STAR-RIS-assisted V2X communication system, the observed states at time $t$, $s_t \in \mathcal{S}$ include wireless channels $\mathbf{h}_i$, amplitude vector of reflection/transmission $\boldsymbol{\beta}_\lambda$, phase shift coefficient vector of reflection/transmission $\boldsymbol{\theta}_\lambda$, digital beamforming vectors for VUEs $\mathbf{p}_\mathrm{V2I}$, and SINR vector of V2V pairs $\boldsymbol{\gamma}_v$, remaining load for transmission $D^r$, and remaining latency $T^r$,  respectively.
 \subsubsection{Actions} The action set consists of all possibles actions that the agent can take in each state. The actions include spectrum allocation $\mathbf{A}$, amplitude vector $\boldsymbol{\beta}_\lambda$ and phase shift coefficient vector $\boldsymbol{\theta}_\lambda$ of reflection/transmission, and transmit power vector from V2V transmitter $\mathbf{p}_\mathrm{V2V}$, respectively. For our system model, each distinct action is defined differently. The amplitude is defined as the incremental amplitude of the current reflection/transmission pattern and can be written as
 \begin{equation}
     \boldsymbol{\beta}_{\lambda}^{t+1} = \boldsymbol{\beta}_{\lambda}^t \odot \triangle \boldsymbol{\beta}_{\lambda}^t,
 \end{equation}
where $\boldsymbol{\beta}_{\lambda}^{t+1}$ is the amplitude vector of reflection/transmission at the $t+1$-th time, $\odot$ is the Hadamard (element-wise) product, $\boldsymbol{\beta}_{\lambda}^t$ is the amplitude vector of reflection/transmission at time $t$, and $\triangle \boldsymbol{\beta}_{\lambda}^t$ is the incremental amplitude of $\boldsymbol{\beta}_{\lambda}^t$. Similarly, the phase shift of the current reflection/transmission pattern can be defined as the incremental phase shift as follows.
 \begin{equation}
     \boldsymbol{\theta}_{\lambda}^{t+1} = \boldsymbol{\theta}_{\lambda}^t \odot \triangle \boldsymbol{\theta}_{\lambda}^t
 \end{equation}
where $\boldsymbol{\theta}_{\lambda}^{t+1}$ is the phase shift vector of reflection/transmission at the $t+1$-th time, $\boldsymbol{\theta}_{\lambda}^t$ is the phase shift vector of reflection/transmission at time $t$, and $\triangle \boldsymbol{\theta}_{\lambda}^t$ is the incremental phase shift of $\boldsymbol{\theta}_{\lambda}^t$, respectively. The spectrum allocation variable is determined as $a_{i,v} \in \{0,1\}$.
The transmit power of V2V transmitter for the $v$-th V2V pair $p_v$ is defined as $p_v \in \{0, (1/L_p-1) P_{\mathrm{V}\_{\max}}, (2/L_p-1) P_{\mathrm{V}\_{\max}}, \dots, P_{\mathrm{V}\_{\max}}\},$ where $L_p$ is the distinct levels of transmit power since we assume discrete power control scheme.
\subsubsection{Transition Function} The transition function $\mathcal{P}(s'|s, a)$ specifies how the environment $s$ evolves over time in response to predicted actions $a$.
\subsubsection{Reward} The goal of DRL is to maximize the reward that represents the desired performance metrics. In our scenario, it is necessary to satisfy the QoS requirements, which are the achievable data rate for VUEs, as well as the reliability and latency requirements for V2V pairs. Therefore, our immediate reward function at time $t$ can be defined as follows.
\begin{align}
&\begin{aligned}
    \mathcal{R}_t = &\sum_{i \in \mathcal{I}}q_1R_i +  \sum_{i \in \mathcal{I}}q_2F\left(R_i - R_{\min}\right) + \\
    &\sum_{v \in \mathcal{V}}q_3F\left(R_v - \frac{D}{T_{\max}}\right) + \sum_{v \in \mathcal{V}}q_4F\left(\gamma_v - \gamma_{\mathrm{ef}}\right),
\end{aligned}
\end{align}
where $F(x)$ is a piece-wise function and can be determined as
\begin{equation}
    F(x) = \left \{ \begin{array}{ll}{P_0} & {\text{when } x \geq 0}, \\ {x,} & {\text {otherwise,}}\end{array}\right.
\end{equation}
where $P_0>0$ is specified as positive constant to signify revenue. The aforementioned MDP problem can be solved using iterative techniques, such as dynamic programming and Q-learning. However, dynamic programming requires complete information about the environment and reward function, which is challenging in highly dynamic network environment. We therefore utilize Q-learning in order to maximize the expected value of the accumulated total reward.

Under given policy $\pi$, the true value of an action $a$ at state $s$ is
\begin{equation}
    Q_\pi(s,a) \equiv \mathbb{E}[\mathcal{R}_1 + \zeta \mathcal{R}_2 + \dots | S_t=s, A_t=a].
\end{equation}
Then, the optimal action values (Q-function) can be calculated as $Q^*(s,a) = \max_\pi Q_\pi(s,a)$. By applying Bellman equation, $Q^*(s,a)$ can be obtained as
\begin{equation}
    Q^*(s,a) = \sum_{s' \in \mathcal{S}}\mathcal{P}(s'|s, a)\left(\mathcal{R}(s,a) + \zeta \max_{a' \in \mathcal{A}} Q^*(s',a')\right).
\end{equation}
Using Q-learning, the estimates of optimal values can be learned, where the optimal action values can be derived from a table lookup. However, since the state of the environment is complex, it is impractical to use a single look-up table to store all Q values, and doing so is tedious.

\subsubsection{DQN} DQN uses deep neural networks to approximate Q-values, which represent the expected cumulative rewards of taking a specific action from a certain state. DQN combines neural networks with Q-learning to learn optimal actions in an environment with rewards and actions. The Q-network denoted as $Q(s,a;\mathbf{w})$ is a deep neural network with weights $\mathbf{w}$ that takes the state $s$ as input and outputs the Q-values for all possible actions $a$. The Q-learning update rule for DQN  relates the Q-values of the current state-action pair to the Q-values of the next state and the immediate reward. The update rule is given by
\begin{align}
&\begin{aligned}
    Q_{t+1}(s,a&;\mathbf{w}) = Q_t(s,a;\mathbf{w}) + \\
    & \alpha \left( \underbrace{\mathcal{R}(s,a) + \zeta \max_{a' \in \mathcal{A}} Q_t(s', a';\mathbf{w}')}_{\mathrm{Target\_DQN}} - Q_t(s,a;\mathbf{w}) \right),
\end{aligned}
\end{align}
where $\alpha \in (0,1]$ is the learning rate to determine the step size of the update. The Q-learning algorithm uses this update rule to iteratively improve the Q-values for each state-action pair, eventually leading to an optimal policy that maximizes the expected cumulative reward. The target network with weights $\mathbf{w}'$ is used to calculate the target Q-values for training to address the issue of unstable target estimates. Still, DQN faces several challenges in their training process, including overestimation of Q-values, instability in learning, and slow convergence. DDQN is a modification of DQN that aims to address these challenges.
\subsubsection{DDQN} Instead of using the target network for both action selection and evaluation, DDQN utilizes two separate neural networks, a primary network, and a target network. The primary network is used for action selection, while the target network is used for action evaluation. This allows DDQN to reduce the overestimation of Q-values, leading to more stable and accurate learning \cite{van2016deep}. The Q-value update for DDQN is given by
\begin{align}
&\begin{aligned}
    & Q_{t+1}(s,a;\mathbf{w}) = Q_t(s,a;\mathbf{w}) + \\
    & \alpha \left( \underbrace{\mathcal{R}(s,a) + \zeta Q(s',\max_{a' \in \mathcal{A}} Q_t(s', a;\mathbf{w});\mathbf{w}')}_{\mathrm{Target\_DDQN}} - Q_t(s,a;\mathbf{w}) \right).
\end{aligned}
\end{align}

\subsubsection{Attention}
Attention is a mechanism used in deep learning models that allows the model to focus on different parts of the input selectively. It provides a way to assign importance weights to different elements of the input based on their relevance or importance for the task at hand. Attention mechanisms have been widely used in natural language processing, computer vision, and other domains to enhance model performance \cite{galassi2020attention, guo2022attention}. Attention mechanisms typically involve three main components. The first component consists of query, key, and value, where the attention mechanism compares a query with a set of key-value pairs. The query represents the current input element for which attention weights are calculated. The key-value pairs represent elements of the input that are being attended to. The second component is attention weights, which represent the importance or relevance of each key with respect to the given query. The third component is the weighted sum or context vector which represents the aggregated information from the values, weighted by their corresponding attention weights. Initially, with query $Q'$, key $K'$, and value $V'$ scaled dot-product attention is computed by applying a softmax function to achieve the weights on the values and can be represented as follows.
\begin{equation}
    \mathrm{Att}(Q', K', V') = \mathrm{softmax}\left(\frac{Q'K'^T}{\sqrt{d_{k'}}}\right),
\end{equation}
where $d_k$ is the input dimension.

\begin{figure}[t]
	\includegraphics[width=\linewidth]{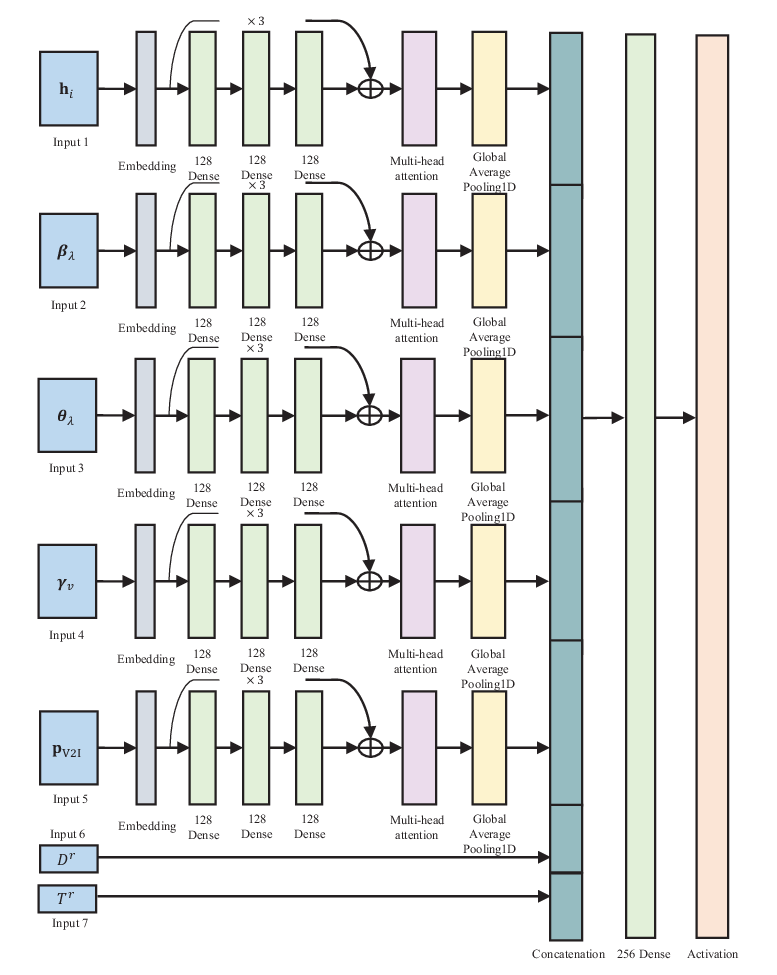}
	\caption{Network architecture of DDQN-attention.}
	\label{na}
\end{figure}

Instead of relying on a single attention mechanism, multi-head attention (MHA) employs multiple parallel attention mechanisms or heads. MHA is an extension of the attention mechanism that enables the model to attend to different aspects of the input simultaneously. Each head learns a different representation or focus of the input, allowing the model to capture diverse patterns and relationships. The use of MHA in models like Transformer has been instrumental in improving the model's ability to capture diverse dependencies and patterns in input sequences \cite{vaswani2017attention}. With a single attention head, averaging inhibits the following
\begin{equation}
    \mathrm{MHA}(Q',K',V') = \mathrm{Concat(Att_1, Att_2, \dots, Att_{\max})}.
\end{equation}
In our network architecture, we apply embeddings to the input states for continuous representation, utilize ResNet to capture complex patterns, incorporate the MHA layer for selective focus on relevant parts of the states, use a dense network to fuse the sub-states obtained from the attention layer, and employ the Swish activation function. The architecture of our network is shown in Fig. \ref{na}.


\subsection{Digital Beamforming Problem}\label{dbp}
For the given spectrum allocation $\mathbf{A}$, amplitude vector of reflection/transmission $\boldsymbol{\beta}_\lambda$, phase shift coefficient of reflection/transmission $\boldsymbol{\theta}_\lambda$, transmit power vector from V2V transmitter $\mathbf{p}_\mathrm{V2V}$ obtained from Section \ref{DA-STAR}, the digital beamforming problem can be represented as
\begin{maxi!}|s|
	{\mathbf{p}_\mathrm{V2I}} {\sum_{i \in \mathcal{I}}R_i}
	{\label{DBOF}}{}\label{DBOFa}
	\addConstraint{R_i \geq R_{\min}, \forall i \in \mathcal{I}}\label{c11}
    \addConstraint{\gamma_v \leq \gamma_{\mathrm{ef}} = \frac{\gamma_0}{\ln{\left(\frac{1}{1-p_0}\right)}}, \forall v \in \mathcal{V}}\label{c33}
    \addConstraint{\mathbf{p}_\mathrm{V2I}^H \mathbf{p}_\mathrm{V2I} \leq P_{\mathrm{I}\_{\max}}}\label{c77},
\end{maxi!}
where (\ref{DBOF}) is non-convex due to the quadratic constraint in (\ref{c77}). Therefore, we relax constraint (\ref{c77}) by using the convexity property of the norm function. Specifically, we replace the quadratic term with the Euclidean norm of $\mathbf{p}_\mathrm{V2I}$. This relaxation results in a convex constraint as follows.
\begin{equation}\label{relnorm}
\norm{\mathbf{p}_\mathrm{V2I}}_2 \leq \sqrt{P_{\mathrm{I}\_{\max}}},
\end{equation}
where $\norm{\mathbf{p}_\mathrm{V2I}}_2$ denoted the Euclidean norm of $\mathbf{p}_\mathrm{V2I}$. Since the Euclidean norm is a convex function, constraint (\ref{relnorm}) represents a convex set in the space of $P_{\mathrm{I}\_{\max}}$. By relaxing the quadratic constraint, the resulting optimization in (\ref{DBOF}) can be rewritten as
\begin{maxi!}|s|
	{\mathbf{p}_\mathrm{V2I}} {\sum_{i \in \mathcal{I}}W_0 \log_{2}\left(1+\frac{|\mathbf{h}_i\mathbf{p}_i|^2}{G_{v}+\sigma^2}\right)}
	{\label{DBOF2}}{}\label{DBOFa2}
	\addConstraint{R_i \geq R_{\min}, \forall i \in \mathcal{I}}\label{c112}
    \addConstraint{\frac{p_v |h_v|^2}{ \sum_{i \in \mathcal{I}} \left(a_{i,v} |\mathbf{h}_i \mathbf{p}_i|^2\right) + \sigma^2} \leq \frac{\gamma_0}{\ln{\left(\frac{1}{1-p_0}\right)}},  \forall v \in \mathcal{V}}\label{c133}
    \addConstraint{\norm{\mathbf{p}_\mathrm{V2I}}_2 \leq \sqrt{P_{\mathrm{I}\_{\max}}}}\label{c772}.
\end{maxi!}
Moreover, as in \cite{yang2021energy}, the slack variable $\boldsymbol{\mu} = [\mu_1, \mu_2, \dots, \mu_I]$ is introduced to guarantee that constraint (\ref{c112}) is always satisfied with equality for the optimal solution, and problem (\ref{DBOF2}) can be rewritten as
\begin{maxi!}|s|[1]
	{\mathbf{p}_\mathrm{V2I}, \boldsymbol{\mu}} {\sum_{i \in \mathcal{I}}W_0 \log_{2}(1+\mu_i)}
	{\label{DBOF3}}{}\label{DBOFa3}
	\addConstraint{\mu_i \leq \frac{|\mathbf{h}_i\mathbf{p}_i|^2}{G_{v}+\sigma^2}, \forall i \in \mathcal{I}}\label{cslack1}
    \addConstraint{\sum_{i \in \mathcal{I}} \left(a_{i,v} |\mathbf{h}_i \mathbf{p}_i|^2\right) + \sigma^2 \geq \frac{p_v |h_v|^2}{\gamma_0} \ln{\left(\frac{1}{1-p_0}\right)},  \forall v \in \mathcal{V}}\label{cslack2}
    \addConstraint{\norm{\mathbf{p}_\mathrm{V2I}}_2 \leq \sqrt{P_{\mathrm{I}\_{\max}}}}\label{cslack3}
    \addConstraint{\mu_i \geq 2^{\frac{R_{\min}}{B}}-1, \forall i \in \mathcal{I}}.\label{cslack4}
\end{maxi!}
The problem in (\ref{DBOF3}) remains non-convex due to constraint (\ref{cslack1}). To address this, a slack variable $\xi_i > 0$ is introduced and constraint (\ref{cslack1}) is reformulated as follows.
\begin{equation}\label{rotation}
    |\mathbf{h}_i\mathbf{p}_i|^2 \geq \xi_i \mu_i.
\end{equation}
Furthermore, the term $|\mathbf{h}_i\mathbf{p}_i|^2$ can be represented as a real value by performing an arbitrary rotation on the digital beamforming vector $\mathbf{p}_i$. Consequently, we obtain $\mathbb{R}(\mathbf{h}_i\mathbf{p}_i) \geq \sqrt{\xi_i \mu_i}$. To transform constraint (\ref{rotation}), we employ the first-order Taylor approximation in place of the concave function $\sqrt{\xi_i \mu_i}$. This yields the following rewritten form for constraint (\ref{rotation}):
\begin{align}\label{taylor}
&\begin{aligned}
    \mathbb{R}(\mathbf{h}_i\mathbf{p}_i) & \geq \sqrt{\xi_i^{(j-1)} \mu_i^{(j-1)}} + \frac{1}{2} \sqrt{\frac{\xi_i^{(j-1)}}{\mu_i^{(j-1)}}}(\mu_i - \mu_i^{(j-1)}) \\
    & + \frac{1}{2}  \sqrt{\frac{\mu_i^{(j-1)}}{\xi_i^{(j-1)}}}(\xi_i - \xi_i^{(j-1)}).
\end{aligned}
\end{align}
Finally, problem (\ref{DBOF3}) can be rewritten as
\begin{maxi!}|s|
	{\mathbf{p}_\mathrm{V2I}, \boldsymbol{\mu}, \boldsymbol{\xi}} {\sum_{i \in \mathcal{I}}W_0 \log_{2}(1+\mu_i)}
	{\label{DBOF4}}{}\label{DBOFa4}
	\addConstraint{(\ref{cslack2}), (\ref{cslack3}), (\ref{cslack4}), (\ref{taylor})}\label{final1}
    \addConstraint{\xi_i \geq 0, \forall i \in \mathcal{I}.}\label{final2}
\end{maxi!}
Problem (\ref{DBOF4}) becomes a convex problem, and can be solved by utilizing CVXPY toolkit in python programming. Our overall proposed algorithm is shown in Algorithm \ref{algo}.

\begin{algorithm}[t] \caption{Combined DDQN-attention and optimization based STAR-RIS-assisted V2X communication system algorithm}\label{algo}
	\begin{algorithmic}[1]
	    \renewcommand{\algorithmicrequire}{\textbf{Input:}}
		\Require Parameters such as discount factor $\xi$, learning rate, mini batch size, replay memory with size $S_{m}$;
		\State \textbf{Initialization} Initialize DDQN-attention with random weights $\mathbf{w}$ and target network weights $\mathbf{w}' = \mathbf{w}$.
        \For{each episode = $1, 2, \dots,$}
            \State Initialize the STAR-RIS-assisted V2X network environment and receive initial observed state
            \For{each time step $t = 1, 2, \dots,$}
                \State Forwards $\mathbf{p}_\mathrm{V2I}$ from problem (\ref{DBOF4}) and other observed state $s_t$ to DDQN-attention.
                \State Observe the states and achieve the state action value $Q_t(s,a;\mathbf{w})$.
                \State Make a copy to obtain target action value function $Q_t(s',a';\mathbf{w}')$.
                \State Based on $\epsilon$ greedy policy, select an action $a_t$.
                \State Obtain reward $\mathcal{R}_t$ and $s_{t+1}.$
                \State Store the tuple $(s_t, a_t, \mathcal{R}_t, s_{t+1})$ into the replay memory.
                \State Randomly select a mini batch of tuples from the replay memory and calculate $\mathrm{Target_{DQN}}$.
                \State Train the DDQN-attention and update its weight $\mathbf{w}$.
                \State For every $S_Q$ steps, update weights of target network as $\mathbf{w}' = \mathbf{w}$.
                \State From the state $s_t$, problem (\ref{DBOF4}) and fed back to the CVXPY to obtain the optimal $\mathbf{p}_\mathrm{V2I}$.
            \EndFor
        \EndFor
    \renewcommand{\algorithmicrequire}{\textbf{Output:}}
	\end{algorithmic}
\end{algorithm}

\subsection{Complexity Analysis}\label{complexity}
According to \cite{al2022reconfigurable} and \cite{yang2021energy}, the complexity of our solution approach can be acquired by each solution method. For joint spectrum allocation, amplitude, phase shift and transmit power problem, the DDQN-attention method is applied, and the complexity of a connected network with $\hat{L}$ layers is denoted as $\mathcal{O}\left(\sum_{\hat{l}}^{\hat{L}}(n_{\hat{l}} n_{\hat{l}-1})\right)$ where $n_{\hat{l}}$ denotes the total number of neurons in layer $\hat{l}$. For the digital beamforming problem, the complexity for the optimization method is $\mathcal{O}(V)$. Therefore, overall computation complexity can be expressed as $\mathcal{O}\left( \bar{T}(\sum_{\hat{l}}^{\hat{L}}(n_{\hat{l}} n_{\hat{l}-1}) + V) \right)$, where $\bar{T}$ represents the number of iterations required to reach convergence.

\section{Performance Evaluation}\label{evaluation}
In this section, we evaluate our STAR-RIS-assisted V2X communication system via numerical results. We consider an area with a road segment of 120 m $\times$ 20 m which consists of two lanes in each direction. The vehicles are equipped with single antennas and the arrival of vehicles is established by the spatial Poisson process. Among the vehicles, the number of active VUEs is 20, and 6 active V2V transmitters are selected randomly. Each V2V transmitter determines a V2V connection with the closest vehicle within its broadcast range. The position of the BS equipped with a uniform linear array (ULA) is $(0,0,10)$ in Cartesian coordinates. The position of STAR-RIS is $(60,10,5)$ which we assume is located on the roadway median, and the number of elements is 32. The simulation parameters can be listed in Table \ref{simtable}.
\begin{table}
	\centering
	\caption{Simulation parameters}
	\begin{tabular}{|l|c|}
		\hline
		\textbf{Parameter}                 & \textbf{Value}    \\
		\hline \hline
		Outage threshold $\gamma_{\mathrm{ef}}$ & 4 dB \\ \hline
        Maximum transmit power budget $P_{\mathrm{I}\_{\max}}$ & 10 dB \\ \hline
		Bandwidth $W$		& 10 GHz \\ \hline
		Noise power $\sigma^2$		& -174 dBm \\ \hline
		Path loss exponent $\delta$                & 4 \\ \hline
		Channel gain at reference distance $\eta$ & -40 dBm \\ \hline
		Rician factor $\hat{R}$                     & 10 \\ \hline
		Discount factor & 0.98 \\ \hline
        Initial exploration & 1 \\ \hline
        Exploration rate & 0.02 \\ \hline
		Learning rate & 0.001 \\ \hline
		Discount factor $\xi$ & 0.9 \\ \hline
		Mini batch size & 4 \\ \hline
		Number of episodes & 1,000 \\ \hline
	\end{tabular}
	\label{simtable}
\end{table}

To compare our proposed STAR-RIS-assisted V2X communication network, we compare with the following benchmark schemes:
\begin{itemize}
    \item \emph{STAR-RIS (Random):} In this scheme, the spectrum allocation matrix $\mathbf{A}$, and the reflection/transmission coefficient of STAR-RIS is random (i.e., random amplitude vector $\boldsymbol{\beta}_\lambda$ and random phase shift vector $\boldsymbol{\theta}_\lambda$). The transmit power vector from V2V transmitter between each V2V pair $\mathbf{p}_{\mathrm{V2V}}$ is solved by applying DDQN-attention algorithm, and digital beamforming vectors $\mathbf{p}_{\mathrm{V2I}}$ is solved by optimization method.
    \item \emph{RIS (Proposed):} In this design, we replace the conventional RIS that supports reflecting only (i.e., $\beta_r = 1, \beta_t = 0$) with STAR-RIS, and implement our proposed algorithm to solve the problem.
    \item \emph{RIS (Random):} In this scheme, the reflection coefficient of conventional RIS is random (i.e., random phase shift vector $\boldsymbol{\theta}_r$). The rest of the control variables are solved by our proposed algorithm.
\end{itemize}

\begin{figure}[t]
	\includegraphics[width=\linewidth]{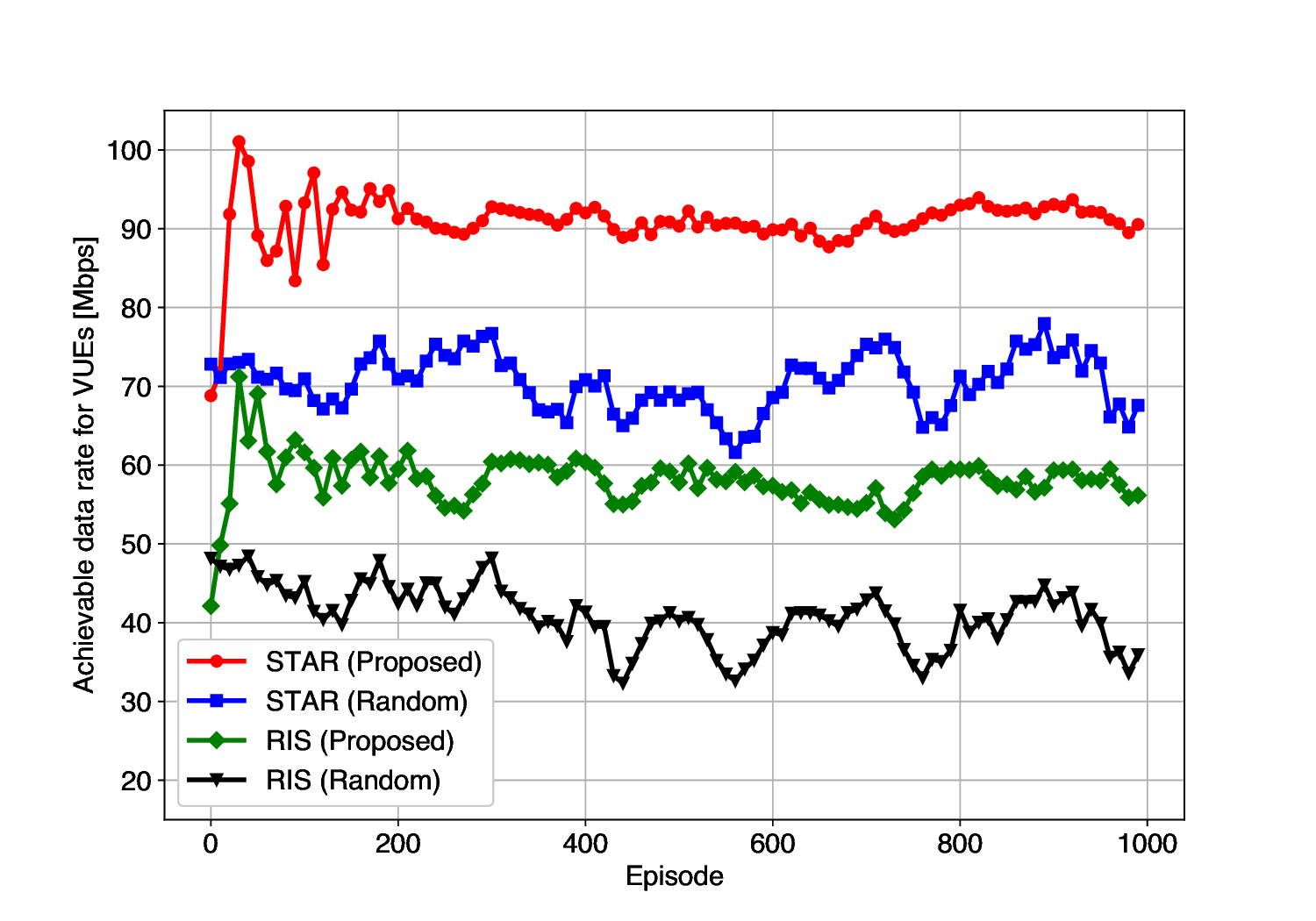}
	\caption{Achievable data rate for VUEs with different benchmark scenarios.}
	\label{starvsris}
\end{figure}
Fig.~\ref{starvsris} illustrates the achievable data rate by all VUEs within our proposed scenario in comparison to the benchmark schemes previously mentioned. It is observed that STAR-RIS-assisted V2X communication networks outperform RIS-assisted V2X communication networks under all circumstances. This superiority can primarily be attributed to the fact that conventional RIS solely supports one side of the road it reflects, while STAR-RIS offers enhanced flexibility by jointly optimizing the amplitude and phase shift coefficients for both reflection and transmission. In addition, our proposed algorithms outperform random schemes, owing to the lack of an optimized performance guarantee in the latter.
\begin{figure}[t]
	\includegraphics[width=\linewidth]{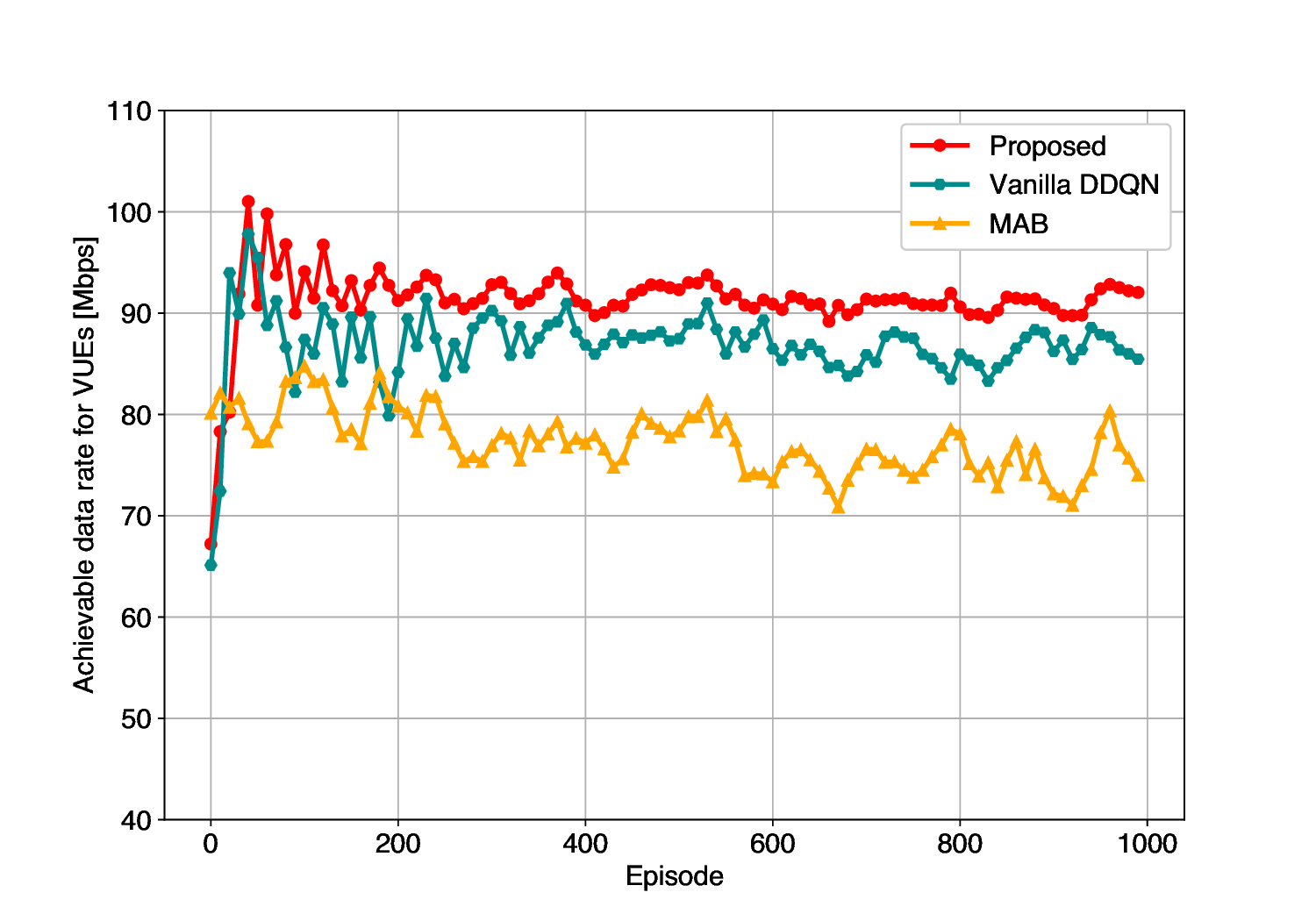}
	\caption{Achievable data rate for VUEs with different benchmark solution algorithms.}
	\label{methods}
\end{figure}
Fig.~\ref{methods} compares the convergence of our proposed DDQN-attention algorithm with vanilla DDQN, where only the deep residual network (ResNet) is applied, and the multi-armed bandit (MAB) algorithm, where the algorithm chooses the arm with the highest estimated reward with the probability 1-$\epsilon$ and selects a random arm with the probability $\epsilon$ \cite{slivkins2019introduction}. As demonstrated, while the performance of our proposed algorithm is marginally superior to that of vanilla DDQN, our proposed algorithm has a faster convergence rate. This is because our proposed algorithm enables the agent to recognize more complex patterns and dependencies in the input. The attention mechanism assists the model in focusing on informative regions of the state and reduces the impact of noisy or irrelevant information. Meanwhile, both DDQN algorithms outperform MAB due to the latter's limited capacity to learn policies and struggle to adapt to changing conditions in complex environments.
\begin{figure}[t]
	\includegraphics[width=\linewidth]{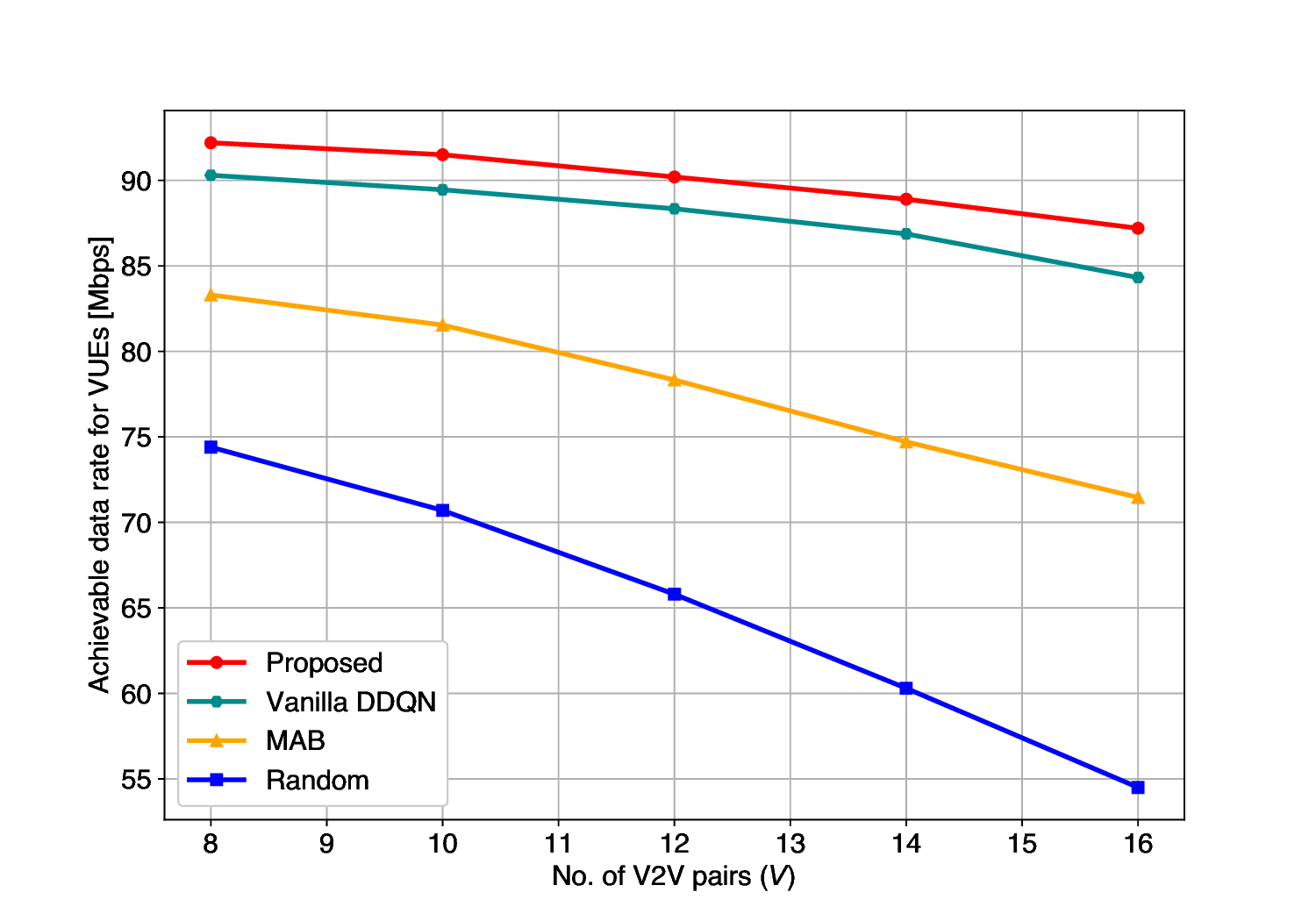}
	\caption{Performance comparison of achievable data rate for VUEs under different number of V2V pairs.}
	\label{vuevsv2v}
\end{figure}
\begin{figure}[t]
	\includegraphics[width=\linewidth]{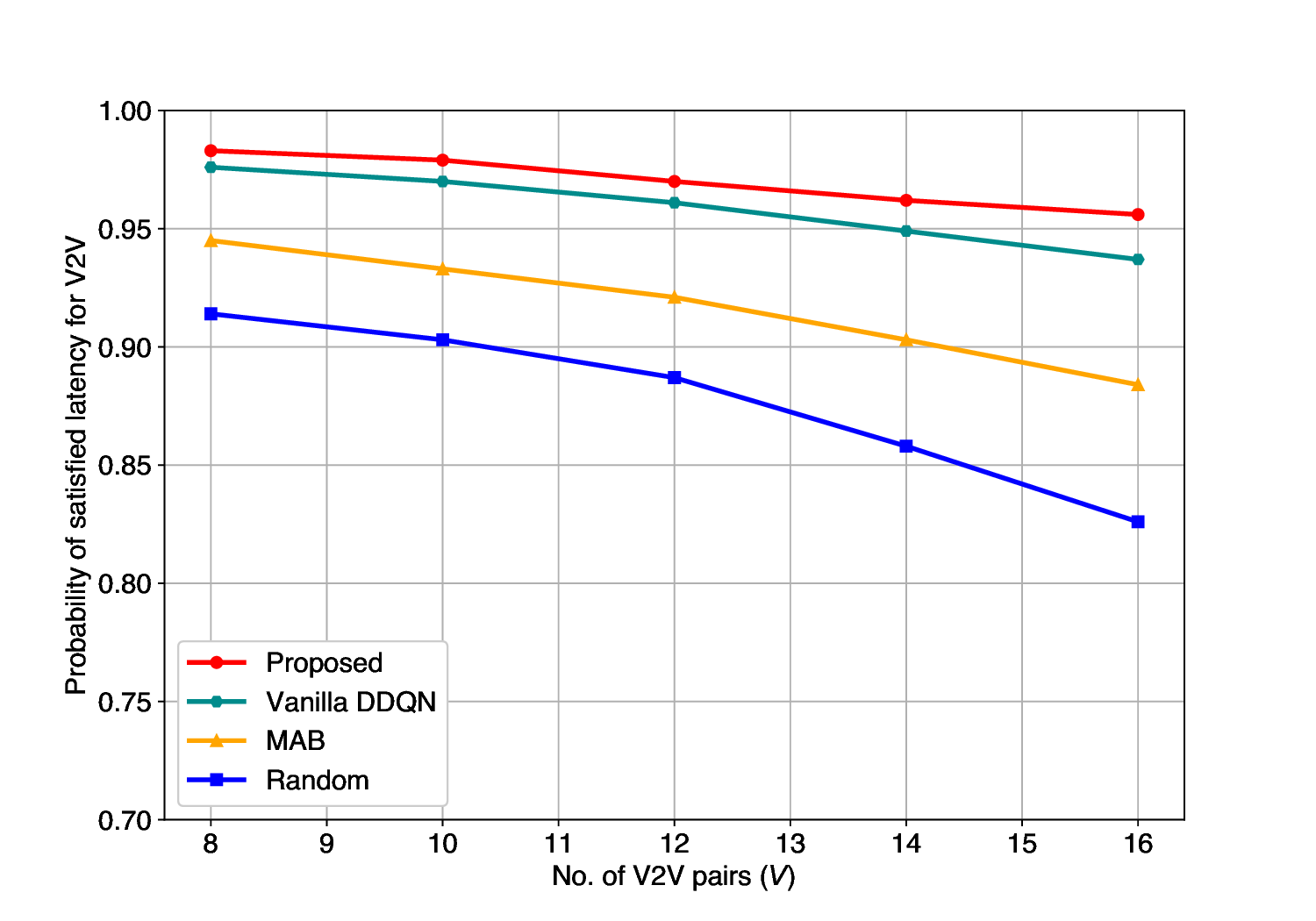}
	\caption{Performance comparison of the probability of satisfied latency for V2V links under different number of V2V pairs.}
	\label{probv2vvsno}
\end{figure}

Fig.~\ref{vuevsv2v} represents the achievable data rate for VUEs with varying numbers of V2V pairs. As depicted, when the number of V2V pairs increases, the obstruction between the BS to STAR-RIS to VUEs links increases, which decreases the achievable data rate for VUEs. Among them, our proposed algorithm outperforms the benchmark schemes since it can detect these V2V links and optimize transmission/reflection coefficients of STAR-RIS and digital beamforming vectors for VUEs. Furthermore, Fig.~\ref{probv2vvsno} represents the probability of the satisfied latency constraint for V2V links. With more V2V pairs, ensuring every vehicle's latency constraint is difficult, thus, reducing the satisfied probability. Our proposed algorithm achieves the highest probability since it can dynamically adjust the transmit power for V2V transmitters and enhance spectrum utilization efficiency for the V2V pairs to reuse the same spectrum with the VUEs.


\begin{figure}[t]
	\includegraphics[width=\linewidth]{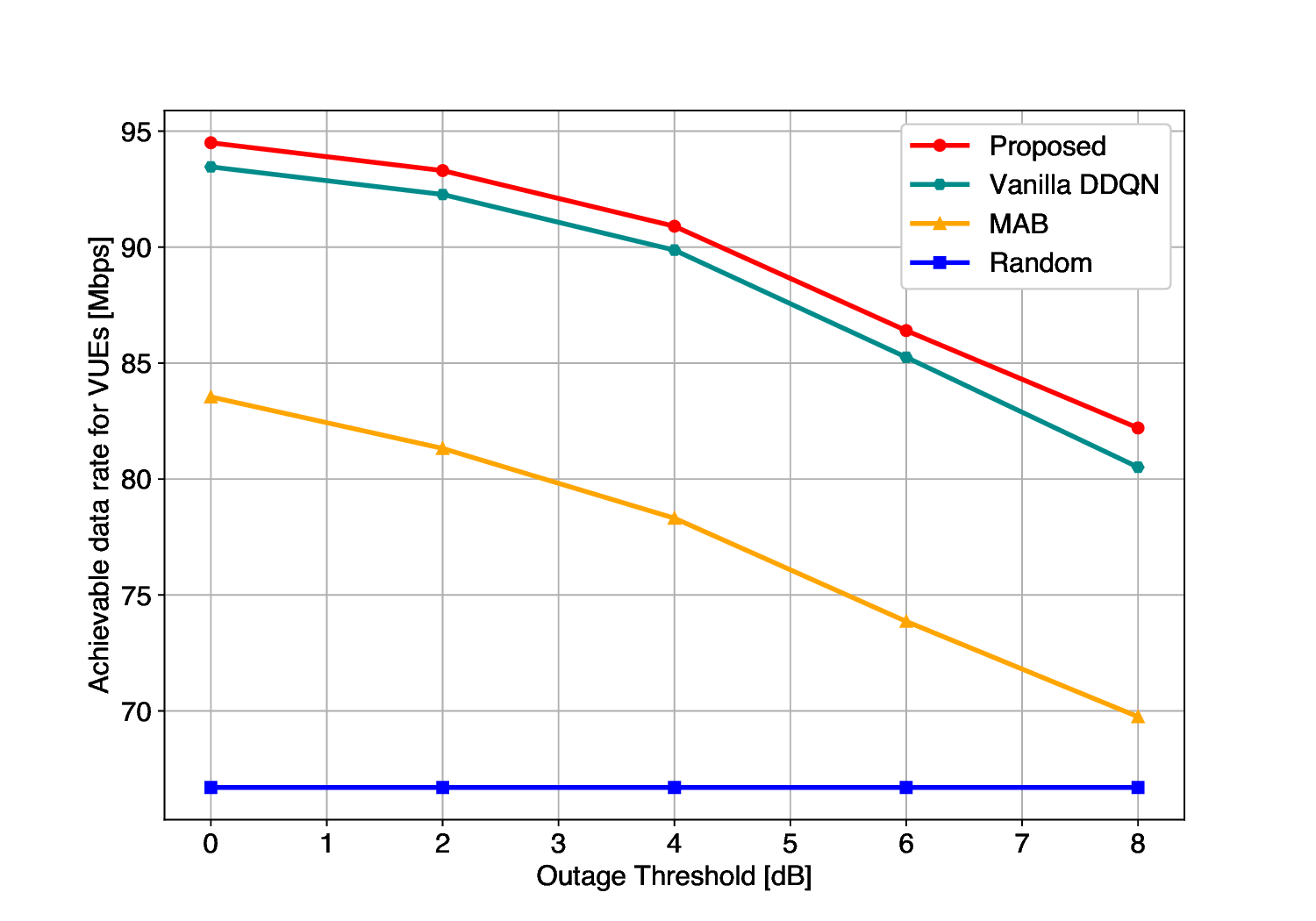}
	\caption{Performance comparison of achievable data rate for VUEs under different outage thresholds.}
	\label{outagev2i}
\end{figure}

\begin{figure}[t]
	\includegraphics[width=\linewidth]{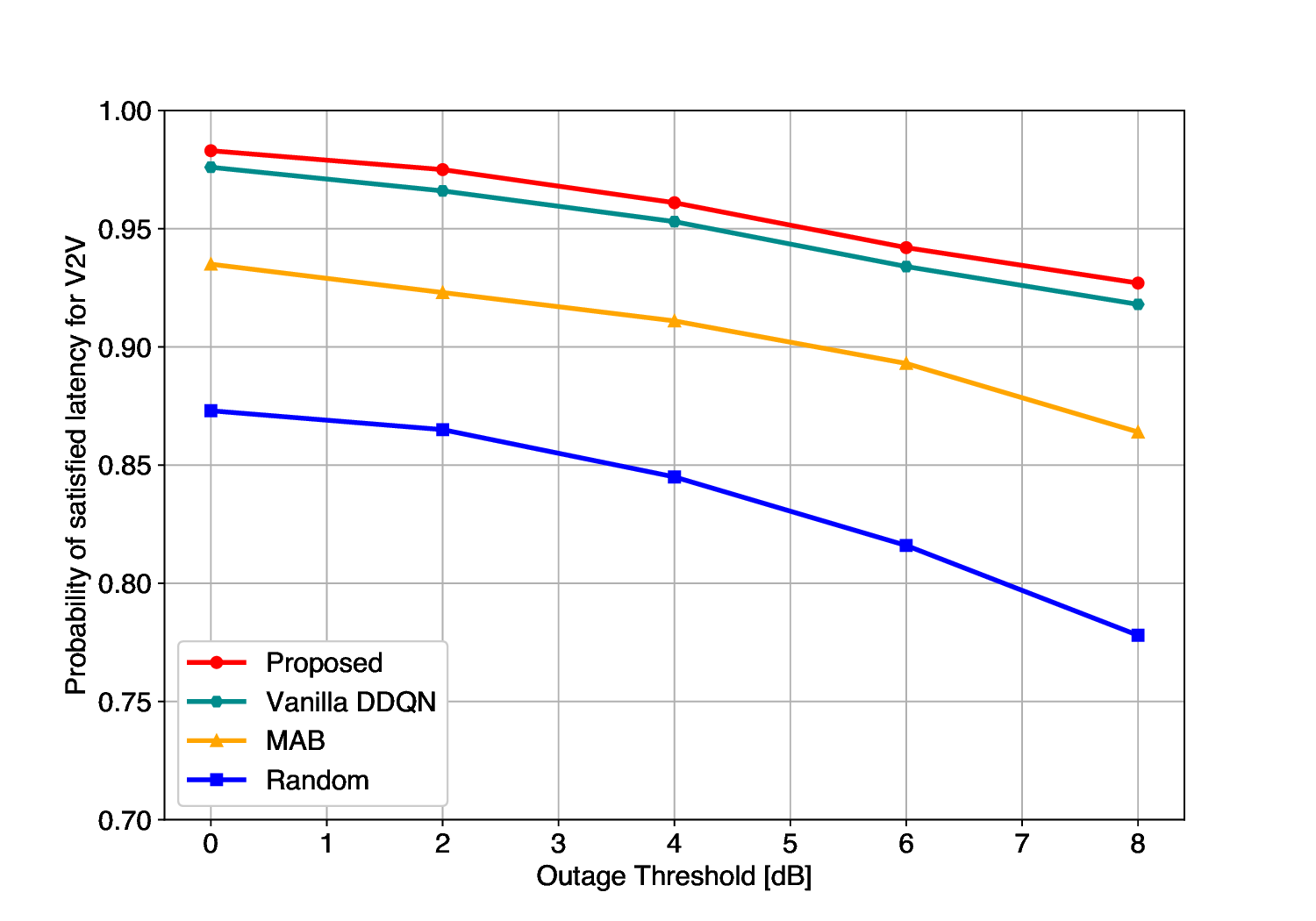}
	\caption{Performance comparison of probability of satisfied latency for V2V links under different outage threshold.}
	\label{outagev2v}
\end{figure}

Moreover, Figs.~\ref{outagev2i} and \ref{outagev2v} compare the achievable data rate for VUEs and the probability of the satisfied latency constraints for V2V links under different outage thresholds. Similarly, with the increment in outage threshold, both the achievable data rate for VUEs and the probability of the satisfied latency decrease. This is because, with a more significant outage threshold, the system allows for a higher probability of experiencing link failures in VUEs and increased tolerance for disruptions which violates the latency constraints in V2V communication. Among them, our proposed DDQN-attention algorithm achieves the minimum decline compared to the benchmark schemes. This is because our algorithm selectively attends to necessary network conditions, such as signal strength and interference levels, which influence the achievable data rate and latency satisfaction.
\begin{figure}[t]
	\includegraphics[width=\linewidth]{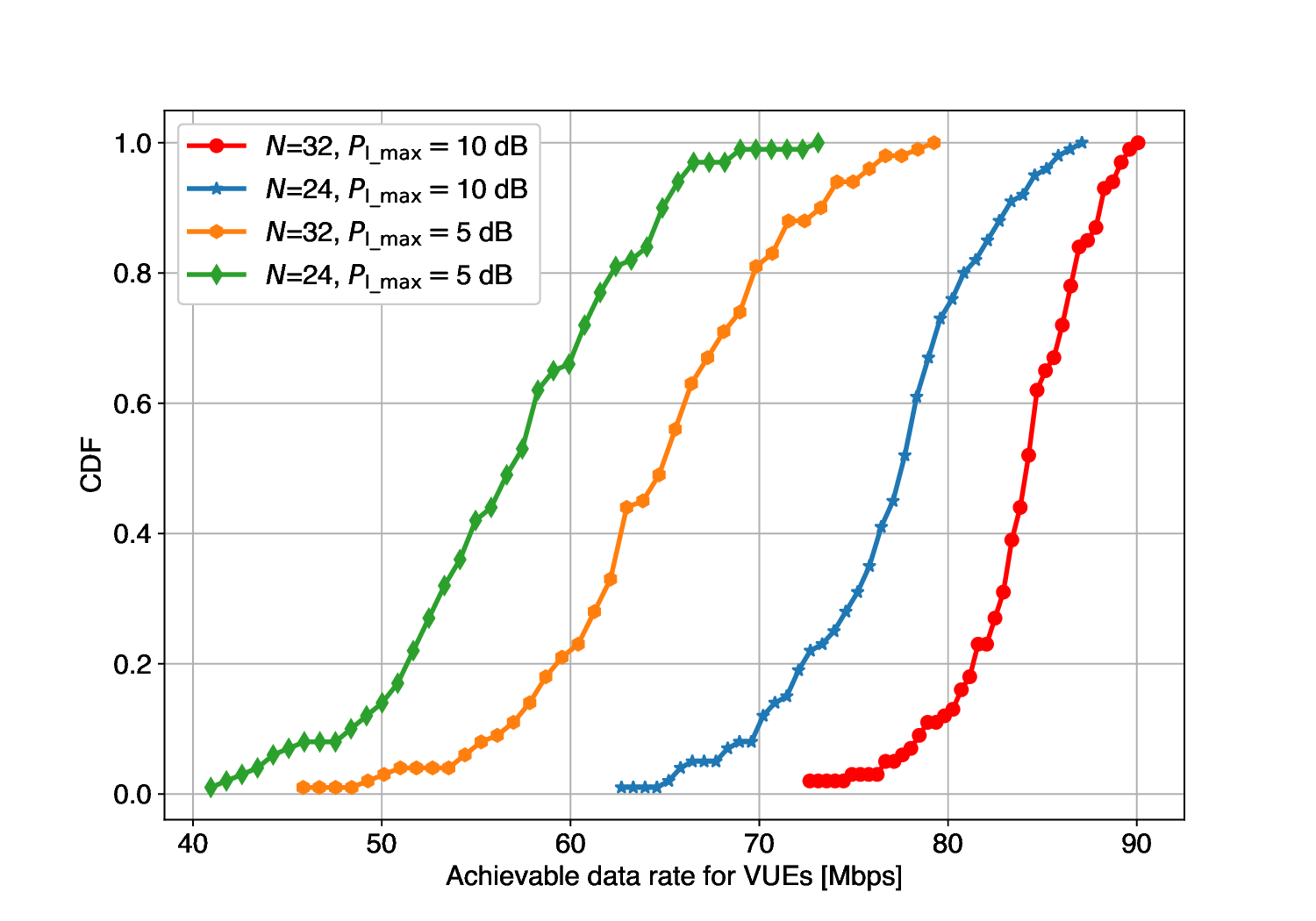}
	\caption{CDF of achievable data rate for VUEs under different number of elements of STAR-RIS and power budgets.}
	\label{cdf}
\end{figure}

Finally, we examine the achievable data rate for VUEs with different elements and power budgets. As shown in Fig. \ref{cdf}, we can observe that when the power budget is low, increasing the number of elements in the system has a significant impact on the data rate for VUEs. This is because the additional elements in the system help to enhance the signal power and direct it toward the intended VUEs. With a higher power budget, the VUEs already receive an adequate level of signal power and quality, and additional elements may contribute to marginal improvements but are not as crucial as in the low power budget scenario.

\section{Conclusion}\label{conclusion}
In this paper, we have studied a STAR-RIS-assisted V2X communication system. To maximize the achievable data rate for VUEs while satisfying the reliability and latency requirements for V2V pairs, we formulated a joint spectrum allocation, amplitudes, and phase shifts of STAR-RIS elements, digital beamforming vectors for VUEs, and transmit power for V2V pairs problem. Since the formulated problem was MINLP and the environment in V2X communications is constantly changing and inconsistent, we separated our problem into two sub-problems, and proposed our combined DDQN-attention and optimization-based approach to solve the problem. Extensive numerical analysis was evaluated to prove that our proposed system outperformed the conventional RIS-assisted V2X communication system and our proposed solution outperformed the vanilla DDQN and MAB algorithms, respectively. The attention enables the model to selectively prioritize on pertinent data, thereby enhancing the performance compared to a conventional network. Furthermore, the results confirm that the STAR-RIS-assisted V2X communication network achieves a better achievable data rate for VUEs compared to conventional RIS due to the better coverage. By employing STAR-RIS, we can not only increase the spectral efficiency for VUEs but also meet the requirements for reliability and latency, which improves the diverse QoS requirements for V2X communications.
\bibliographystyle{IEEEtran}
\bibliography{mybib}









\end{document}